\documentstyle[pre,aps,multicol]{revtex}
\input epsf

\begin{document}

\preprint{FIMAT-7/95 {\bf /} MA/UC3M/08/95}

\draft

\title{Extended states and dynamical localization in the random-dimer
model}

\author{F.\ Dom\'{\i}nguez-Adame$^{\dag}$}

\address{Departamento de F\'{\i}sica de Materiales,
Facultad de F\'{\i}sicas, Universidad Complutense,
E-28040 Madrid, Spain}

\author{Angel S\'{a}nchez$^{\ddag}$ and Enrique Diez$^{*}$}

\address{Departamento de Matem\'aticas, Escuela Polit\'ecnica Superior,
Universidad Carlos III, E-28911 Legan\'{e}s, Madrid, Spain}

\date{\today}

\maketitle

\begin{abstract}

We study quantum diffusion of wavepackets in one-dimensional random
binary subject to an applied electric field.  We consider three
different cases: Periodic, random, and random dimer (paired) lattices.
We analyze the spatial extent of electronic wavepackets by means of the
time-dependent inverse participatio ratio.  We show that the delocalized
states recently found in random dimer lattices become spatially
localized under the action of the applied field (dynamical localization)
but wavepackets are much less localized than in purely random lattices.
We conclude that the resonant tunneling effects causing delocalization
play an important role even in the presence of the electric field.

\end{abstract}


\begin{multicols}{2}

\narrowtext

\section{Introduction}

The existence of one-dimensional (1D) disordered lattices with a number
of extended electronic states large enough to contribute to transport
properties in a relevant fashion has been undoubtedly established during
this decade.  The interest on this problem arose from the pioneering
works of Flores \cite{Flo} and Dunlap, Wu and Philips \cite{Dun}, which
stimulated a considerable effort devoted to understand these
delocalization phenomena \cite{Datta1,PRB1,Sou,D,Hilke,ind,Hein}.  The
common feature of the models studied so far is that they consist of a
host (discrete or continuum) system where defects are placed randomly
although their distribution exhibit spatial correlation.  This spatial
correlation is usually introduced by imposing that impurities appear
always in pairs (dimers) or in more complicated grouping schemes.
Specifically, results for 1D random models with paired disorder, i.e.,
with defects forming dimers, which exhibit delocalization have been put
on solid theoretical grounds \cite{rusos}.  Interestingly, spatial
correlations in 1D random systems lead to new and unexpected phenomena
not only in electronic systems but also in the case of quantum
ferromagnets \cite{spin}, Frenkel excitons \cite{PRBF}, classical
vibrations \cite{PRB2} and excitations \cite{PRBC}.  It thus becomes
clear that the study of random systems with correlated disorder is of
interest in a wide range of physical problems, and that, ultimately,
such a line of research may lead to the development of a variety of new
devices and applications.

In Ref.\ \cite{rusos}, it was shown that delocalized electronic states
arise in spite of the inherent disorder due to resonant phenomena taking
place at dimers which, in turn, lead to a transmission coefficient of
different segments forming the lattice close to unity, no matter what
the length of the segment is.  The transmission coefficient is exactly
unity for the resonant energy at a single dimer defect and, most
important, is very close to unity for electron energies near the
resonant one, as demonstrated by perturbative and numerical
calculations.  Once the existence of these bands of extended states is
put forward and the reasons for its appearance are understood in the
isolated, non-interacting model, an interesting question immediately
arises, namely what is the effect of external perturbations on the
delocalized states?  In particular, since we are going to concern
ourselves with electronic states, the first perturbation that has to be
studied is an applied electric field; since applied electric fields lead
to localization even in periodic lattices, one should expect that
delocalized states in random correlated systems might also be spatially
localized.  But the key question is to elucidate whether the physical
mechanisms giving rise to delocalization in unperturbed systems are of
relevance in the presence of the electric field or, on the contrary,
they are immaterial at all.  The answer to this question is not trivial:
Competition between quantum coherence due to correlated disorder and the
loss of quantum coherence due to the misalignement of local electronic
levels under the action of the field will be the main mechanism
governing this system, and the prevailing factor among these two is
difficult to foresee.

In this letter we present a first study in the above direction.  We
consider the problem of quantum diffusion of wavepackets initially
localized in random-dimer models (RDMs), as introduced in
Ref.~\onlinecite{Dun}, under the action of a uniform electric field.
The way we carry out such an analysis is by comparing electronic
amplitudes in three different binary systems, namely periodic, unpaired
disordered lattices and paired disordered ones.  The study of periodic
systems will allow us to establish the main features of dynamical
localization in periodic binary systems.  This is required for a better
understanding of wavepacket dynamics when an amount of randomness is
introduced in the system.  To get an estimation of the spreading of the
wavepacket as a function of time we will use the time-dependent inverse
participatio ratio (IPR).  By means of this quantity, we will be able
to show below that, although all states in random dimer models become
localized under the action of the electric field, they acquire an
spatial structure much more extended that their counterparts in the
purely random lattice.

\section{Model}

The unperturbed random-dimer model is described by the following
1D tight-binding Hamiltonian \cite{Dun}
\begin{equation}
{\cal H}_0=\sum_n\>E_nc_n^{\dag}c_n + V\sum_n\> (c_{n+1}^{\dag}c_n +
c_n^{\dag} c_{n+1}).
\label{H0}
\end{equation}
Here $c_n$ and $c_n^{\dag}$ are electron annihilation and creation
operators in the site representation.  The hopping matrix element $V$ is
assumed to be constant over the whole lattice, whereas on-site energies
$E_n$ can only take on two values $E_A$ and $E_B$, with the additional
constraint that $E_B$ are assigned at random to pairs of lattices sites.
In Ref.\ \cite{Dun}, it was found that for $|E_A-E_B|<2|V|$, an
initially localized wavepacket becomes delocalized and its mean-square
displacement grows in time as $t^{3/2}$ (super-diffusion).  For
$|E_A-E_B|=2|V|$ the mean-square displacement behaves asymptotically as
$t$ (diffusion).  Otherwise the wavepacket remains localized.

As we mentioned above, we are interested in quantum diffusion of
wavepackets under an applied electric field.  In particular, we
investigate the time behaviour for different sets of constituent
parameter $E_A$, $E_B$, and $V$.  The perturbed Hamiltonian is written
as follows
\begin{equation}
{\cal H}={\cal H}_0-F\sum_n\>nc_n^{\dag}c_n,
\label{H}
\end{equation}
where $F$ is the electric field (we use units such that $e=\hbar=1$ in
the rest of the paper).  In order to solve the corresponding
Sch\"odinger equation we express the wave function in terms of localized
Wannier states.  In doing so, it can be seen that the time-dependent
amplitudes $\psi_n(t)$ satisfy the following equation
\begin{equation}
i\,{d\phantom{t}\over dt}\psi_n(t)
=(E_n-Fn)\psi_n+V(\psi_{n-1}+\psi_{n+1}).
\label{psi}
\end{equation}

This equation is solved numerically with the initial condition
$\psi_n(0)=\delta_{n\,n_0}$, where $n_0$ denotes the initial site, using
an implicit (Crank-Nicholson) integration scheme.  Such a procedure
preserves the normalization condition, which has been used at every time
step to test the accuracy of results.  Once the solution is found, the
wavepacket dynamics can be characterized by means of the time-dependent
IPR, defined as follows \cite{Evangelou}
\begin{equation}
\mbox{IPR\,}(t)=\sum_n\>|\psi_n(t)|^4
\label{ipr}
\end{equation}
Usually the value of the IPR is a good estimation of the spatial extent
of electronic states.  Delocalized states are expected to present small
IPR (in the ballistic limit, without applied field, it vanishes as
$t^{-1}$), while localized states have larger IPR (in the limit of
strong localization should be unity whenever the electron is localized
at a single site).

\section{Results and discussions}

In the present work we study three different binary systems: periodic
(ABABAB $\ldots$), unpaired and paired disordered lattices, with a given
fraction $c$ of B-sites.  For definiteness, we take $E_A=0$ and $V=-1$,
whereas $E_B=1,2,3$ to include super-diffusive, diffusive and localized
states of the unperturbed RDM. Although the local environment for
different initial sites $n_0$ is different, we have checked that our
main conclusions remain valid and only finer details of plots change
upon setting several values of $n_0$ and different realizations of
disorder.

\begin{figure}
\setlength{\epsfxsize}{6.5cm}
\centerline{\mbox{\epsffile{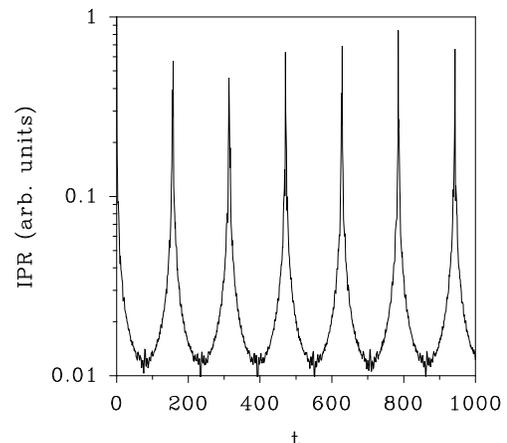}}}
\caption{Inverse participatio ratio as function of time in a binary
periodic lattice with $E_A=0$, $E_B=1$, and $V=-1$, for $F=0.02$.}
\label{fig1}
\end{figure}

We begin by studying the periodic lattice as a way to verify that we are
carrying out our computations correctly.  Figure~\ref{fig1} presents the
results for $F=0.02$ in a binary periodic lattice with $E_A=0$, $E_B=1$,
and $V=-1$.  Similar results are obtained for other values of the
on-site energy $E_B$ and electric field $F$.  The IPR displays marked
peaks at times $t_k=k\tau$ ($k=0,1\ldots$), where $\tau=156.7$ for this
value of the electric field.  The value of maxima at $t_k$ is very close
to unity, indicating that the wavepacket is strongly localized around
the initial site $n_0$ (dynamical localization).  This is consequence of
the so called Bloch oscillations, where the wavepacket oscillates in
time with period $\tau_B = 2\pi\hbar/FL$, $L$ being the lattice period
(see, e.g., Ref~\onlinecite{Dignam}).  In our case $L=2$ and, therefore,
the corresponding period will be $\tau_B=157.1$, in perfect agreement
with that obtained from Fig.~\ref{fig1} We can therefore trust the
results we are going to obtain for the random lattices.

\begin{figure}
\setlength{\epsfxsize}{6.8cm}
\centerline{\mbox{\epsffile{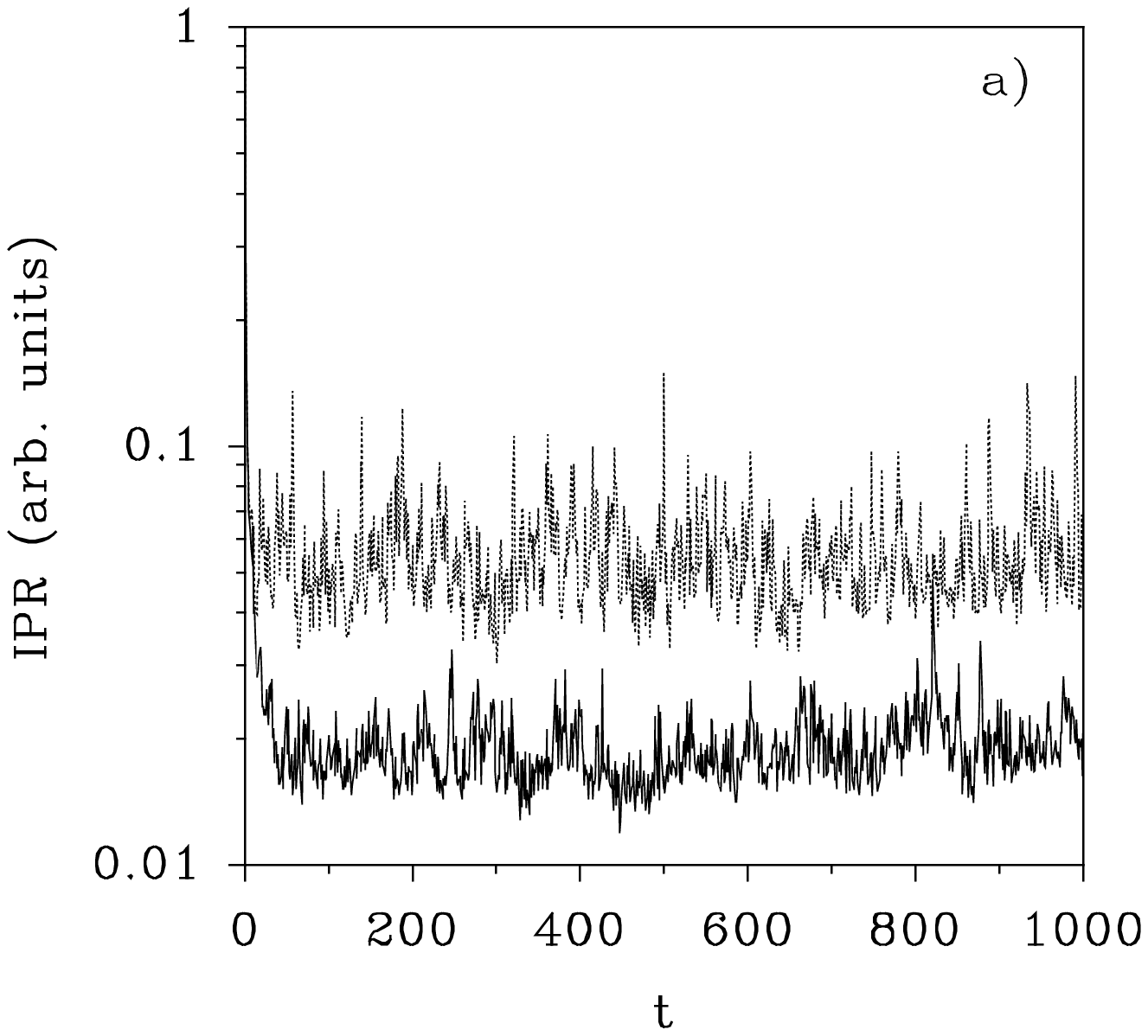}}}
\setlength{\epsfxsize}{6.8cm}
\centerline{\mbox{\epsffile{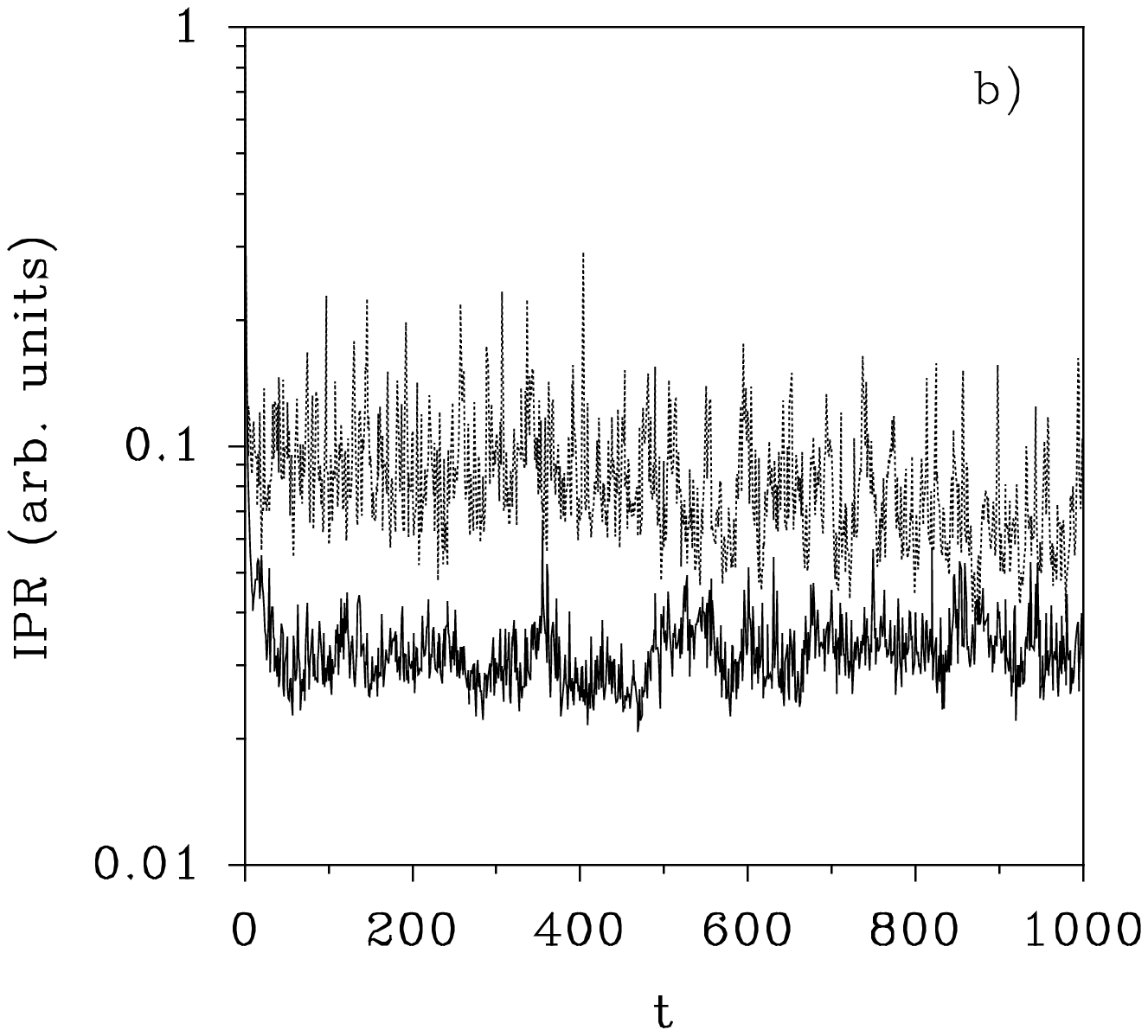}}}
\setlength{\epsfxsize}{6.8cm}
\centerline{\mbox{\epsffile{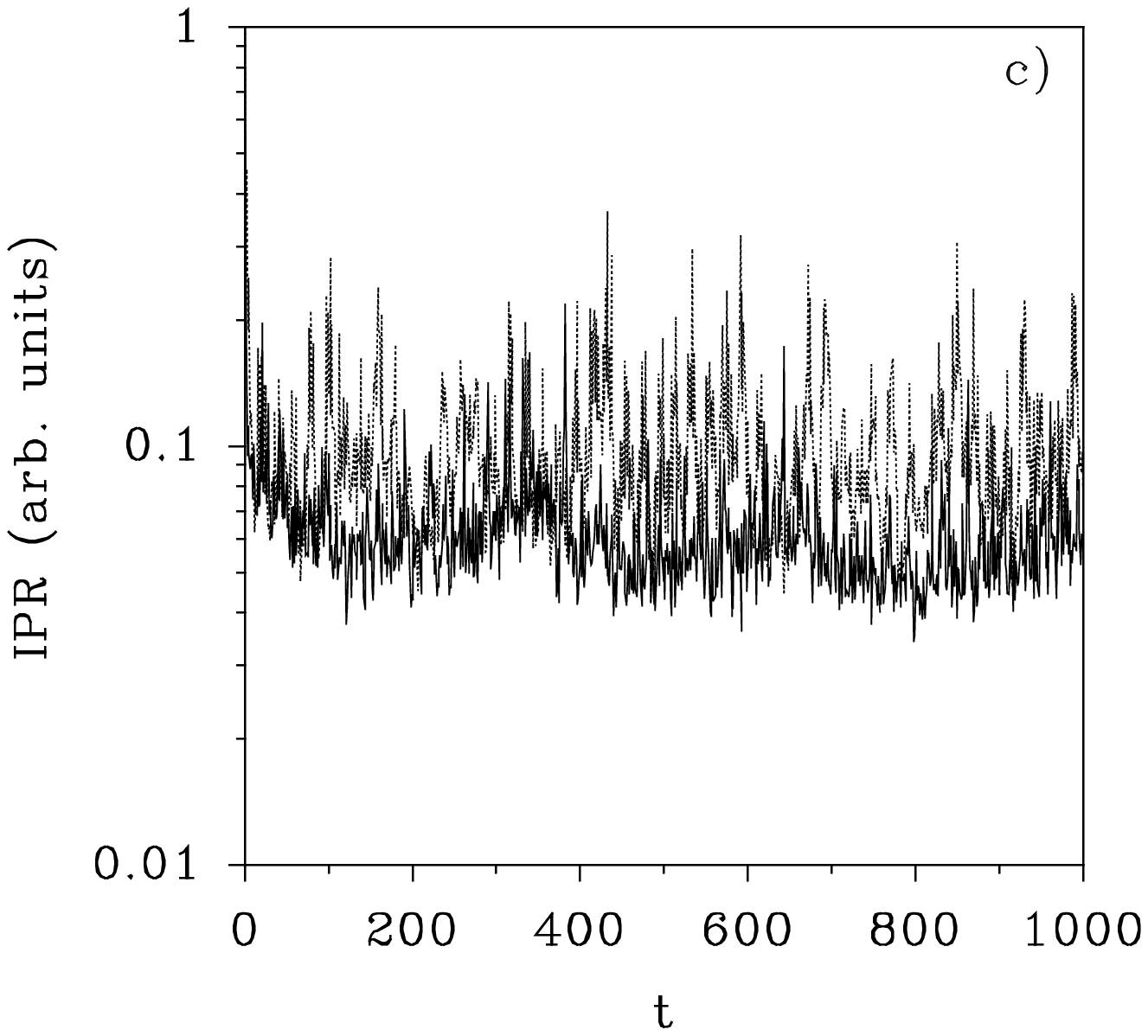}}}
\caption{Inverse participatio ratio as function of time in a binary
random lattices with $E_A=0$, and $V=-1$, for $F=0.02$ with (a)
$E_B=1$, (b) 2, and (c) 3.  Solid and dotted lines correspond to paired
and unpaired disordered lattices, respectively.}
\label{fig2}
\end{figure}

Results corresponding to both kinds of disordered lattices are shown in
Fig.~\ref{fig2}, for $E_A=0$, $V=-1$, $F=0.02$, and three different
values of the on-site energy $E_B$.  The concentration of B-sites is
$c=0.2$ in all cases.  We can observe that Bloch oscillations are
completely absent in random lattices.  This fact can be explained by the
absence of translational invariance and, consequently, by scattering of
electrons with the lattice, which destroys the quantum coherence
required to observe such phenomenon.  In both kind of random lattices
the IPR presents strong fluctuations at small time scales, but it can be
observed that its average value over larger times is constant.  Such
small fluctuations depend on the particular realization of the disorder
and on the initial position of the wavepacket.  However, for a given
concentration $c$, the mean value depends only on the electric field and
on the hopping matrix element (the larger the electric field or the
hopping matrix element, the closer to unity the IPR).

This far, we have summarized the common features of states of both
random lattices.  It is now the moment to comment the main differences
between paired and unpaired lattices.  When $|E_A-E_B|<2|V|$, i.e.
whenever the defect energy lies within the allowed band
[Fig.~\ref{fig2}(a)], the mean value of the IPR is smaller for paired
lattices, meaning that the wavepacket spreads over larger portions of
the system.  From this plot, it can be appreciated that the difference
between both IPR values is about an order of magnitude, and hence the
spatial extent of wavefunctions in paired and unpaired lattices will
largely differ.  Thus, when the unperturbed ($F=0$) paired lattice
supports extended states, the resulting dynamical localization under the
action of the electric field is much {\em less} effective than in
unpaired lattices.  It is therefore reasonable to expect that the
transport properties of the two systems will also exhibit specific
features: For instance, the hopping conductivity has to be much larger
in the dimer lattice than in the random lattice, due to the increased
tunneling probability between neighboring localized states.  A smaller
degree of localization in the dimer lattice is also observed, although
to a somewhat lesser extent, in the critical situation $|E_A-E_B| =
2|V|$ [Fig.~\ref{fig2}(b)].  On the contrary, the dynamics in both
lattices is almost identical whenever the unperturbed lattice only
supports localized states [Fig.~\ref{fig2}(c)].

A better understanding of this result is achieved if one considers that
the (initially strongly localized) wavepacket is the combination of
plane waves in a continuous band \cite{Katsanos}.  Since the energy
spectrum of the paired disordered lattices presents a band of extended
states, the lattice behaves as a selective electronic filter, and those
components whose wavenumber belongs to this band can propagate over
larger distances, producing a larger spreading of the resulting
wavepacket.  The observation of this behaviour, as we have reported, is
therefore a clear consequence of the fact that the unperturbed lattice
supports extended states.  Finally, the absence of Bloch oscillations in
paired disordered lattices indicates that their extended states are no
longer Bloch states.  Bloch states are characterized by a complete
quantum coherence with a perfectly defined phase.  This is not the case
in the RDM, where electronic states increment its phase by a factor of
$\pi$ whenever they pass over a dimer defect \cite{Dun,PRB1}, and the
position of each dimer defect is in any case a random variable.

\section{Conclusion}

We have studied quantum diffusion of wavepackets driven by an applied
external field in periodic and random (unpaired and paired) binary
lattices.  The spatial degree of localization of wavepackets initially
localized at a single lattice site has been properly described by means
of the time-dependent IPR. In binary periodic lattices we have confirmed
the dynamical localization under electric fields as well as Bloch
oscillations, for which the wavepacket oscillates in time with a
well-defined period proportional to the inverse of the electric field.
Quantum dynamics in disordered lattices also exhibits dynamical
localization although it turns out to be much more intricated: In
particular, no evidence of Bloch oscillations (regular behavior) is
observed.  What is most important for the purposes of the present work,
we have determined that dynamical localization is less effective in
paired disordered lattices than in unpaired ones, provided that the
energy of defects lies within the band of extended states.  Thus, it can
be concluded that extended states can spread over larger segments of the
lattice, giving rise to a smaller localization of the wavepacket in the
presence of the electric field.  Therefore, the resonant tunneling
effects causing delocalization plays an important role, even in the
presence of the applied field.

The results we have reported in this letter provide another piece of
evidence supporting the true extended nature of states near the resonant
energy in the random dimer model.  From plots in Figs.~\ref{fig1} and
\ref{fig2}(a), one can observe that the value of the IPR for the random
dimer model is about the minimum. of the Bloch oscillations of the
periodic lattice.  This is to be compared with the ramdom case, whose
IPR is close to half of the IPR of the periodic lattice.  It is then
clear that electric-field-localized states in the random dimer model are
much closer to those of the periodic lattice than to the purely random
system.  In view of this, we envisage that the transport properties of
random dimer lattices under electric fields will also be close to those
of periodic lattices, this being an experimentally verifiable,
qualitative prediction.  To conclude, we mention that another question
stemming from this work is whether the same behavior will be found in
more realistic models such as the continuum random dimer model
\cite{PRB1} or the square well model \cite{PRBF}.  These models have
already given rise to quantitative predictions of effects that should be
observed in real devices and therefore they are very appealing in order
to find physically relevant consequences of dynamical localization in
dimer models.  Work in this direction is in progress.

\acknowledgments

This work has been supported by CICYT (Spain) under project MAT95-0325.

\end{multicols}

\end{document}